\documentclass[journal]{IEEEtran}
\usepackage[T1]{fontenc}
\usepackage{ifpdf}
\usepackage{cite}
\usepackage[pdftex]{graphicx}
\usepackage{amsmath}
\usepackage{amssymb}
\usepackage{bm}
\usepackage{array}
\usepackage{fixltx2e}
\usepackage{stfloats}

\usepackage[hidelinks]{hyperref}

\begin{document}
\title{Dynamic Energy-Efficient Power Allocation in Multibeam Satellite Systems}

\author{Christos~N.~Efrem, and~Athanasios~D.~Panagopoulos,~\IEEEmembership{Senior~Member,~IEEE}
\thanks{C. N. Efrem and A. D. Panagopoulos are with the School of Electrical and Computer Engineering, National Technical University of Athens, 15780 Athens, Greece (e-mails: chefr@central.ntua.gr, thpanag@ece.ntua.gr).

This article has been accepted for publication in \textit{IEEE Wireless Communications Letters}, DOI: 10.1109/LWC.2019.2949277.  Copyright \textcopyright \ 2019 IEEE. Personal use is permitted, but republication/redistribution requires IEEE permission. See \url{http://www.ieee.org/publications_standards/publications/rights/index.html} for more information.
}}


\maketitle

\begin{abstract}
Power consumption is a major limitation in the downlink of multibeam satellite systems, since it has a significant impact on the mass and lifetime of the satellite. In this context, we study a new energy-aware power allocation problem that aims to jointly minimize the unmet system capacity (USC) and total radiated power by means of multi-objective optimization. First, we transform the original nonconvex-nondifferentiable problem into an equivalent nonconvex-differentiable form by introducing auxiliary variables. Subsequently, we design a successive convex approximation (SCA) algorithm in order to attain a stationary point with reasonable complexity. Due to its fast convergence, this algorithm is suitable for dynamic resource allocation in emerging on-board processing technologies. In addition, we formally prove a new result about the complexity of the SCA method, in the general case, that complements the existing literature where the complexity of this method is only numerically analyzed.  

\end{abstract}


\IEEEpeerreviewmaketitle

\section{Introduction}

\IEEEPARstart{M}{ultibeam} satellite systems (MSS) provide flexibility and efficient exploitation of the available resources in order to satisfy the (potentially asymmetric) traffic demand of users. Due to the fact that the satellite power is quite limited, resource allocation mechanisms should take into consideration not only the co-channel interference (CCI), but also the satellite power consumption in the downlink transmission.

The joint problem of routing and power allocation in MSS is examined in \cite{Neely}, using Lyapunov stability theory. Moreover, the studies \cite{Choi_09} and \cite{Choi_05} deal with several resource allocation problems in MSS with and without CCI, respectively. In \cite{Destounis}, a dynamic power allocation algorithm is proposed exploiting a rain attenuation stochastic model. A comparison between non-orthogonal frequency reuse (NOFR) and beam-hopping (BH) systems is presented in \cite{Lei}, where various capacity optimization schemes are reported. Furthermore, linear and nonlinear precoding techniques are investigated in \cite{Christopoulos} and \cite{Zheng}. 

Unlike previous works, a multi-objective approach that minimizes the USC together with the satellite power consumption is presented in \cite{Aravanis}. In particular, a two-stage optimization is proposed to attain a set of Pareto optimal solutions using metaheuristics. However, these algorithms do not provide any optimality guarantee, and their performance is heavily affected by the optimization parameters. Besides, although this method is suitable for offline power allocation, it is rather inappropriate for online/real-time power allocation since it requires a lot of computation time to find nearly-optimal solutions.

In this letter, we introduce a new performance metric, which has not been systematically studied so far, including both the USC and total power consumption. This is in contrast to the majority of recent studies that solely minimize either the former or the latter objective. Moreover, we develop an optimization algorithm which always converges and, assuming appropriate constraint qualifications, achieves a stationary point (first-order optimality guarantee) with relatively low complexity. In addition, numerical results show that the algorithm performance is almost independent of the initialization point. Consequently, the proposed algorithm can be used in dynamic wireless environments where the resource allocation should be decided in a very short time. Finally, a formal proof about the complexity of the SCA method is also given. 

The rest of this study is organized as follows. In Section II, the optimization problem is formulated and then transformed into an equivalent differentiable form. Afterwards, based on the SCA method, we design an energy-efficient power allocation algorithm in Section III. The performance of this algorithm is analyzed through simulations in Section IV, and some conclusions are provided in Section V. 

\section{Problem Formulation and Transformation}
Consider a multibeam satellite system with a geostationary satellite using $N$ beams ($\mathcal{N} = \{ 1,2, \ldots ,N\}$) and 
$K$ subcarriers (SCs) of bandwidth ${B_{SC}}$ ($\mathcal{K} = \{ 1,2, \ldots ,K\}$). For notation simplicity and without loss of generality, it is assumed that: 1) the total bandwidth, ${B_{tot}} = K {B_{SC}}$, is reused by all beams, i.e., the frequency reuse factor is equal to 1 (worst-case scenario), and 2) during a specific time slot, each beam serves only one user within its coverage area (user $i$ is served by the ${i^{th}}$ satellite beam, $\forall i \in \mathcal{N}$). Moreover, we focus on the downlink (data transmission from the satellite to users) considering ideal, without noise and interference, feeder links between the gateways and the satellite. 

The signal to interference-and-noise ratio (SINR) of the ${i^{th}}$ user ($i \in \mathcal{N}$) on the ${k^{th}}$ SC ($k \in \mathcal{K}$) is expressed by: \linebreak $\gamma _i^{[k]} = {{g_{i,i}^{[k]}p_i^{[k]}} \mathord{\left/ {\vphantom {{g_{i,i}^{[k]}p_i^{[k]}} {\left( {\sum\limits_{j \in \mathcal{N}\backslash i} {g_{j,i}^{[k]}p_j^{[k]}}  + \sigma _{i,k}^2} \right)}}} \right. \kern-\nulldelimiterspace} {\left( {\sum\limits_{j \in \mathcal{N}\backslash i} {g_{j,i}^{[k]}p_j^{[k]}}  + \sigma _{i,k}^2} \right)}}$, where $p_j^{[k]}$ is the transmit power of the ${j^{th}}$ satellite beam, $\sigma _{i,k}^2$ is the thermal noise power of the  ${i^{th}}$ user, and $g_{j,i}^{[k]}$ is the channel power gain between the ${j^{th}}$ satellite beam and the ${i^{th}}$ user, all over the ${k^{th}}$ SC. More precisely, $g_{j,i}^{[k]}$ includes free-space path loss (FSPL), rain attenuation, transmit antenna gain of satellite beam as well as receive antenna gain of user. For the sake of convenience, the transmit power vector is denoted by ${\mathbf{p}} = \left[ {{{\mathbf{p}}^{[1]}},{{\mathbf{p}}^{[2]}}, \ldots ,{{\mathbf{p}}^{[K]}}} \right]$, where ${{\mathbf{p}}^{[k]}} = \left[ {p_1^{[k]},p_2^{[k]}, \ldots ,p_N^{[k]}} \right]$, $\forall k \in \mathcal{K}$. In addition, the USC \cite{Anzalchi} is defined by:
\begin{equation}
USC({\mathbf{p}}) = \sum\limits_{i \in \mathcal{N}} {\max \left( {C_i^{req} - {C_i}({\mathbf{p}}),0} \right)}
\end{equation}
where $C_i^{req}$ and ${C_i}({\mathbf{p}}) = {B_{SC}}\sum\limits_{k \in \mathcal{K}} {{{\log }_2}\left( {1 + \gamma _i^{[k]}} \right)}$ are the ${i^{th}}$ user's requested and offered capacity (in bps), respectively\footnote{In case of \textit{adaptive coding and modulation (ACM)}, the offered capacity can be approximated by ${C_i^{ACM}}({\mathbf{p}}) \approx {B_{SC}}\sum\limits_{k \in \mathcal{K}} {{{\log }_2}\left( {1 + \zeta \gamma _i^{[k]}} \right)}$ without altering the methodology, where $\zeta \in (0,1)$ is obtained through curve fitting (offered capacity versus SINR).}. Moreover, the total radiated power is given by:
\begin{equation}
{P_{tot}}({\mathbf{p}}) = \sum\limits_{i \in \mathcal{N}} {\sum\limits_{k \in \mathcal{K}} {p_i^{[k]}} }
\end{equation}

Focusing on the multi-objective optimization, we study the following nonconvex minimization problem: 
\begin{equation} \label{original_problem}
\mathop {\min }\limits_{{\mathbf{p}} \in Z} \quad f({\mathbf{p}}) = USC({\mathbf{p}}) + w{\kern 1pt} {P_{tot}}({\mathbf{p}})
\end{equation}
with convex feasible set $Z = \{ {\mathbf{p}} \in \mathbb{R}_ + ^{NK}:\; \sum\limits_{k \in \mathcal{K}} {p_i^{[k]}}  \leqslant P_{{\kern 1pt} i}^{\max }, \allowbreak \;\forall i \in \mathcal{N} \;\text{and}\; \sum\limits_{i \in \mathcal{N}} {\sum\limits_{k \in \mathcal{K}} {p_i^{[k]}} }  \leqslant P_{{\kern 1pt} tot}^{\max }\}$, where $P_{{\kern 1pt} i}^{\max }$ is the maximum transmit power of the ${i^{th}}$ satellite beam, and $P_{{\kern 1pt} tot}^{\max }$ is the maximum total radiated power of the satellite\footnote{It is possible to have additional \textit{minimum-capacity constraints for each user} (${C_i}({\mathbf{p}}) \geqslant C_i^{min}$, $\forall i \in \mathcal{N}$) in order to increase the system availability (the methodology remains the same).}. The fixed/predefined weight $w{\kern 1pt}  \in [0, + \infty )$ is measured in bps/W, and expresses the priority of the total radiated power with respect to USC. Consequently, a trade-off between the USC and total power consumption (which is proportional to the total radiated power) can be achieved for a specific value of $w$. In particular, $w = 0$ corresponds to USC minimization. Moreover, it can be proved that problem \eqref{original_problem} is NP-hard by following similar arguments as in \cite{Aravanis}. Nevertheless, as will be seen later, we can obtain a stationary point of the equivalent differentiable problem with reasonable complexity. 

Afterwards, by applying the transformation ${\mathbf{p}} = {2^{\mathbf{y}}}$ ($p_i^{[k]} = {2^{y_i^{[k]}}}$, $\forall i \in \mathcal{N}$, $k \in \mathcal{K}$), where ${\mathbf{y}} = \left[ {{{\mathbf{y}}^{[1]}},{{\mathbf{y}}^{[2]}}, \ldots ,{{\mathbf{y}}^{[K]}}} \right]$ with ${{\mathbf{y}}^{[k]}} = \left[ {y_1^{[k]},y_2^{[k]}, \ldots ,y_N^{[k]}} \right]$, $\forall k \in \mathcal{K}$, we obtain the equivalent nonconvex problem: 
\begin{equation} \label{transformed_problem}
\mathop {\min }\limits_{{\mathbf{y}} \in S} \quad f({2^{\mathbf{y}}}) = USC({2^{\mathbf{y}}}) + w{\kern 1pt} {P_{tot}}({2^{\mathbf{y}}})
\end{equation}
with convex feasible set $S = \{ {\mathbf{y}} \in {\mathbb{R}^{NK}}:\; \sum\limits_{k \in \mathcal{K}} {{2^{y_i^{[k]}}}}  \leqslant P_{{\kern 1pt} i}^{\max }, \allowbreak \;\forall i \in \mathcal{N} \;\text{and}\; \sum\limits_{i \in \mathcal{N}} {\sum\limits_{k \in \mathcal{K}} {{2^{y_i^{[k]}}}} }  \leqslant P_{{\kern 1pt} tot}^{\max }\}$. Notice that the above transformation reduces the number of constraints by $NK$ (lower complexity), since ${\mathbf{p}} \in \mathbb{R}_ + ^{NK}$ becomes ${\mathbf{y}} \in {\mathbb{R}^{NK}}$.

Finally, in order to remove the non-differentiability of the objective function, we rewrite problem \eqref{transformed_problem} in its \textit{equivalent epigraph-form} \cite{Boyd} using the auxiliary variable \linebreak ${\mathbf{t}} = \left[ {{t_1},{t_2}, \ldots ,{t_N}} \right]$:
\begin{equation} \label{equivalent_problem}
\mathop {\min }\limits_{({\mathbf{y}},{\mathbf{t}}) \in \Omega } \quad F({\mathbf{y}},{\mathbf{t}}) = \sum\limits_{i \in \mathcal{N}} {{t_i}}  + w \sum\limits_{i \in \mathcal{N}} {\sum\limits_{k \in \mathcal{K}} {{2^{y_i^{[k]}}}} }
\end{equation}
with nonconvex feasible set $\Omega  = \{ ({\mathbf{y}},{\mathbf{t}}) \in {\mathbb{R}^{NK + N}}:\; {t_i} \geqslant 0, \allowbreak \;{t_i} \geqslant C_i^{req} - {C_i}({2^{\mathbf{y}}}), \;\forall i \in \mathcal{N} \;\text{and}\; {\mathbf{y}} \in S\}$. Observe that the new objective $F({\mathbf{y}},{\mathbf{t}})$ is convex now, and the first two constraints in $\Omega$ are equivalent to ${t_i} \geqslant \max \left( {C_i^{req} - {C_i}({2^{\mathbf{y}}}),0} \right)$, $\forall i \in \mathcal{N}$. Furthermore, problem \eqref{equivalent_problem} is equivalent to problem \eqref{transformed_problem} in the following sense: $({\mathbf{y}},{\mathbf{t}})$ is a global optimum of \eqref{equivalent_problem} if and only if ${\mathbf{y}}$ is a global optimum of \eqref{transformed_problem} and \linebreak ${t_i} = \max \left( {C_i^{req} - {C_i}({2^{\mathbf{y}}}),0} \right)$, $\forall i \in \mathcal{N}$. 

\section{Energy-Efficient Power Allocation}
Subsequently, we utilize the mathematical tool of SCA (refer to the Appendix) in order to tackle problem \eqref{equivalent_problem} with relatively low complexity. Firstly, the offered capacity can be written as follows: ${C_i}({2^{\mathbf{y}}}) = {B_{SC}}\sum\limits_{k \in \mathcal{K}} {\left[ {\varphi _i^{[k]}\left( {{{\mathbf{y}}^{[k]}}} \right) - \vartheta _i^{[k]}\left( {{{\mathbf{y}}^{[k]}}} \right)} \right]}$, where $\varphi _i^{[k]}\left( {{{\mathbf{y}}^{[k]}}} \right)$ and $\vartheta _i^{[k]}\left( {{{\mathbf{y}}^{[k]}}} \right)$ are convex functions given by (note that the log-sum-exp function is convex \cite{Boyd}):
\begin{equation}
\varphi _i^{[k]}\left( {{{\mathbf{y}}^{[k]}}} \right) = {\log _2}\left( {\sum\limits_{j \in \mathcal{N}} {g_{j,i}^{[k]}{2^{y_j^{[k]}}}}  + \sigma _{i,k}^2} \right)
\end{equation}
\begin{equation}
\vartheta _i^{[k]}\left( {{{\mathbf{y}}^{[k]}}} \right) = {\log _2}\left( {\sum\limits_{j \in \mathcal{N}\backslash i} {g_{j,i}^{[k]}{2^{y_j^{[k]}}}}  + \sigma _{i,k}^2} \right)
\end{equation}

Now, for a given approximation point ${\mathbf{\bar y}} \in {\mathbb{R}^{NK}}$, we can construct the next convex minimization problem: 
\begin{equation} \label{convex_problem}
\mathop {\min }\limits_{({\mathbf{y}},{\mathbf{t}}) \in \Theta ({\mathbf{\bar y}})} \quad F({\mathbf{y}},{\mathbf{t}}) = \sum\limits_{i \in \mathcal{N}} {{t_i}}  + w \sum\limits_{i \in \mathcal{N}} {\sum\limits_{k \in \mathcal{K}} {{2^{y_i^{[k]}}}} }
\end{equation}
with convex feasible set $\Theta ({\mathbf{\bar y}}) = \{ ({\mathbf{y}},{\mathbf{t}}) \in {\mathbb{R}^{NK + N}}:\; {t_i} \geqslant 0, \allowbreak \;{t_i} \geqslant C_i^{req} - \widetilde {{C_i}}({\mathbf{y}},{\mathbf{\bar y}}),\;\forall i \in \mathcal{N} \;\text{and}\; {\mathbf{y}} \in S\}$, where:
\begin{equation}
\widetilde {{C_i}}({\mathbf{y}},{\mathbf{\bar y}}) = {B_{SC}}\sum\limits_{k \in \mathcal{K}} {\left[ {\widetilde {\varphi _i^{[k]}}\left( {{{\mathbf{y}}^{[k]}},{{{\mathbf{\bar y}}}^{[k]}}} \right) - \vartheta _i^{[k]}\left( {{{\mathbf{y}}^{[k]}}} \right)} \right]}
\end{equation}
\begin{equation}
\widetilde {\varphi _i^{[k]}}\left( {{{\mathbf{y}}^{[k]}},{{{\mathbf{\bar y}}}^{[k]}}} \right) = \varphi _i^{[k]}\left( {{{{\mathbf{\bar y}}}^{[k]}}} \right) + \nabla \varphi _i^{[k]}\left( {{{{\mathbf{\bar y}}}^{[k]}}} \right) \cdot {\left( {{{\mathbf{y}}^{[k]}} - {{{\mathbf{\bar y}}}^{[k]}}} \right)^T}
\end{equation}
Observe that $\widetilde {{C_i}}({\mathbf{y}},{\mathbf{\bar y}})$ is a concave function of ${\mathbf{y}}$. In addition, the elements of $\nabla \varphi _i^{[k]}\left( {{{{\mathbf{\bar y}}}^{[k]}}} \right)$ are given by:
\begin{equation}
\frac{{\partial \varphi _i^{[k]}\left( {{{{\mathbf{\bar y}}}^{[k]}}} \right)}}{{\partial y_l^{[k]}}} = \frac{{g_{l,i}^{[k]}{2^{\bar y_l^{[k]}}}}}{{\sum\limits_{j \in \mathcal{N}} {g_{j,i}^{[k]}{2^{\bar y_j^{[k]}}}}  + \sigma _{i,k}^2}},\quad \forall l \in \mathcal{N}
\end{equation}

\begin{table}[!t]
\centering
\begin{tabular}{l}
\hline
\textbf{Algorithm 1.} Energy-Efficient Power Allocation
\\ \hline
1: Select a starting point ${\mathbf{p}} \in Z$, and a tolerance $\epsilon > 0$ \\
2: Set $\ell  = 0$, ${\mathbf{y}} = {\log _2}({\mathbf{p}})$, ${t_i} = \max \left( {C_i^{req} - {C_i}({\mathbf{p}}),0} \right)$, $\forall i \in \mathcal{N}$ \\
~~~\,and ${F_0} = F({\mathbf{y}},{\mathbf{t}})$ \\
3: \textbf{repeat} \\                                                                                                                                            
4:~~~Solve the convex minimization problem \eqref{convex_problem} with approximation \\
~~~\space\enspace point ${\mathbf{\bar y}} = {\mathbf{y}}$ in order to achieve a global optimum $({{\mathbf{y}}^ * },{{\mathbf{t}}^ * })$ \\
5:~~~Set $\ell  = \ell  + 1$, ${\mathbf{y}} = {{\mathbf{y}}^ * }$, ${\mathbf{t}} = {{\mathbf{t}}^ * }$, ${\mathbf{p}} = {2^{\mathbf{y}}}$ and ${F_\ell } = F({\mathbf{y}},{\mathbf{t}})$ \\
6: \textbf{until}                                                                                                                                              $\left| {{F_\ell } - {F_{\ell  - 1}}} \right| \leqslant \epsilon \left| {{F_{\ell  - 1}}} \right| $ \\ \hline
\end{tabular}
\vspace{-2mm}
\end{table}

Algorithm 1 presents an iterative process based on the SCA method. In particular, we provide the next proposition which readily follows from Theorems 1 and 2 in the Appendix. Note that the number of variables and constraints of problem \eqref{convex_problem} is polynomial in  $N$ and $K$ ($NK + N$ and $3N + 1$, respectively).

\vspace{1.5mm}
\noindent
\textbf{Proposition 1.} \textit{Algorithm 1 generates a monotonically decreasing sequence ${\left\{ {{F_\ell }} \right\}_{\ell  \geqslant 0}}$ (i.e., ${F_{\ell  + 1}} \leqslant {F_{\ell }}$) and converges to a finite value $L$ ($\mathop {\lim }\limits_{\ell  \to \infty } {F_{\ell }} = L >  - \infty$). Moreover, assuming suitable constraint qualifications, $L = \mathop {\lim }\limits_{\ell  \to \infty } {F_{\ell }} = F\left( {{\mathbf{\hat y}},{\mathbf{\hat t}}} \right)$ for some stationary point $\left( {{\mathbf{\hat y}},{\mathbf{\hat t}}} \right)$ of problem \eqref{equivalent_problem}. Finally, the complexity of Algorithm 1 is $\mathcal{O}\left( {\left( {{\xi \mathord{\left/{\vphantom {\xi }} \right.\kern-\nulldelimiterspace} } {\epsilon}} \right)h(N,K)} \right)$, where $\xi  = {{{F_0}} \mathord{\left/{\vphantom {{{F_0}} {{F_ * }}}} \right.\kern-\nulldelimiterspace} {{F_ * }}} \geqslant 1$, with ${{F_ * }}$ being the globally minimum objective value of problem \eqref{equivalent_problem}, and $h(N,K)$ is the complexity of the convex problem \eqref{convex_problem} which is polynomial in $N$ and $K$.}

\section{Numerical Simulations and Discussion}
In this section, we examine a MSS with the parameters given in Table I. Unless otherwise specified, the tolerance and the starting point of Algorithm 1 are selected as $\epsilon = {10^{ - 3}}$ and ${\mathbf{p}} = \left( {{{P_{{\kern 1pt} tot}^{\max }} \mathord{\left/{\vphantom {{P_{{\kern 1pt} tot}^{\max }} {(NK)}}} \right.\kern-\nulldelimiterspace} {(NK)}}} \right){{\mathbf{1}}_{1 \times NK}}$, where ${{\mathbf{1}}_{1 \times NK}}$ is the all-ones $1 \times NK$ vector. As concerns the requested capacities of the users, we have assumed an asymmetric traffic distribution according to the linear model: $C_i^{req} = r \, i$, $\forall i \in \mathcal{N}$, where $r$ is the traffic slope measured in bps. Furthermore, each satellite beam antenna has the following radiation pattern \cite{Christopoulos}, \cite{Aravanis}: $G(\theta ) = {G_{\max }}{\left( {\frac{{{J_1}(u)}}{{2u}} + 36\frac{{{J_3}(u)}}{{{u^3}}}} \right)^2}$, where $\theta$ is the angle between the corresponding beam center and the user location with respect to the satellite, ${G_{\max }}$ is the maximum satellite beam antenna gain ($G(0) = {G_{\max }}$), $u = 2.07123\tfrac{{\sin (\theta )}}{{\sin ({\theta _{3\text{dB}}})}}$ with ${\theta _{3\text{dB}}}$ the 3-dB angle ($G({\theta _{3\text{dB}}}) = {{{G_{\max }}} \mathord{\left/{\vphantom {{{G_{\max }}} 2}} \right.\kern-\nulldelimiterspace} 2}$), and ${J_1}(u)$, ${J_3}(u)$ are respectively the first and third order Bessel functions of the first kind.

\begin{table}[!t]
\caption{System Parameters}
\centering
\begin{tabular}{|c|c|}
\hline
\textbf{Parameter} & \textbf{Value} \\  \hline
Beam radius & 150 km \\ \hline
Frequency band & Ka (20 GHz) \\ \hline
Number of beams and SCs & $N = 7$, $K = 4$ \\ \hline
Subcarrier bandwidth (${B_{SC}}$) & 125 MHz \\ \hline
Thermal noise power ($\sigma _{i,k}^2 = {\sigma ^2}$, $\forall i \in \mathcal{N}$, $k \in \mathcal{K}$) & $-$124 dBW \\ \hline
Maximum beam power ($P_{{\kern 1pt} i}^{\max } = {P_{\max }}$, $\forall i \in \mathcal{N}$) & 100 W \\ \hline
Maximum total power ($P_{{\kern 1pt} tot}^{\max }$) & 500 W \\ \hline
Free-space path loss & 210 dB \\ \hline
Rain attenuation mean and standard deviation & 2.6 dB, 1.63 dB \\ \hline
User antenna gain & 41.7 dBi \\ \hline
Maximum satellite beam antenna gain (${G_{\max }}$) & 52 dBi \\ \hline
3-dB angle (${\theta _{3\text{dB}}}$) & ${0.2^ \circ }$ \\ \hline
\end{tabular}
\vspace{-2mm}
\end{table}

\begin{figure}[!t]
\centering
\includegraphics[width=3.35in]{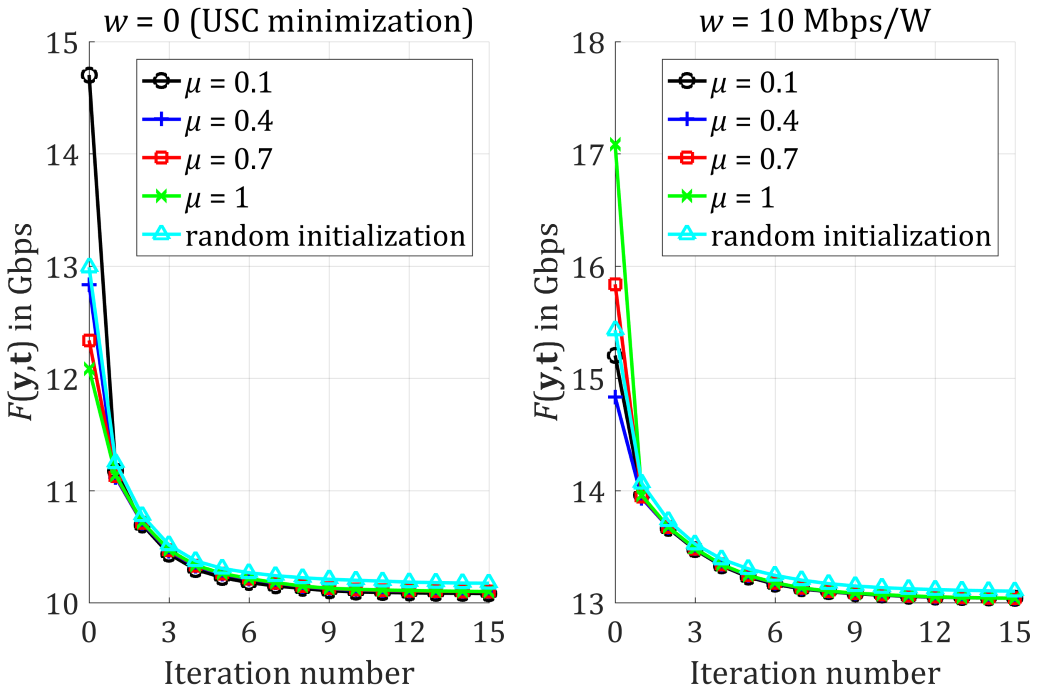} 
\caption{Convergence of Algorithm 1 for $r = 0.7\,\text{Gbps}$, and starting point ${\bf{p}} = \mu \left( P_{{\kern 1pt} tot}^{\max }/(NK) \right){{\bf{1}}_{1 \times NK}}$ or random initialization.}
\label{Fig1}
\vspace{-2mm}
\end{figure}

All graphs, except for Fig. 3, present statistical averages derived from 200 independent Monte Carlo simulations, where each user is uniformly distributed within its beam coverage area. For the sake of comparison, we have used a conventional scheme, namely, uniform power allocation (UPA), where $p_{i,UPA}^{[k]} = {{P_{{\kern 1pt} tot}^{\max }} \mathord{\left/{\vphantom {{P_{{\kern 1pt} tot}^{\max }} {(NK)}}} \right.\kern-\nulldelimiterspace} {(NK)}}$, $\forall i \in \mathcal{N}$, $k \in \mathcal{K}$.

Firstly, we investigate the convergence speed of the proposed algorithm for $w = 0,\,10\,\text{Mbps/W}$ and different starting points. As shown in Fig. 1, Algorithm 1 achieves nearly the same convergence rate and final objective value regardless of the starting point. Given the tolerance $\epsilon = {10^{ - 3}}$, the proposed algorithm requires about 10 iterations to converge for both values of $w$ and for all the starting points under consideration.

\begin{figure}[!t]
\centering
\includegraphics[width=3.35in]{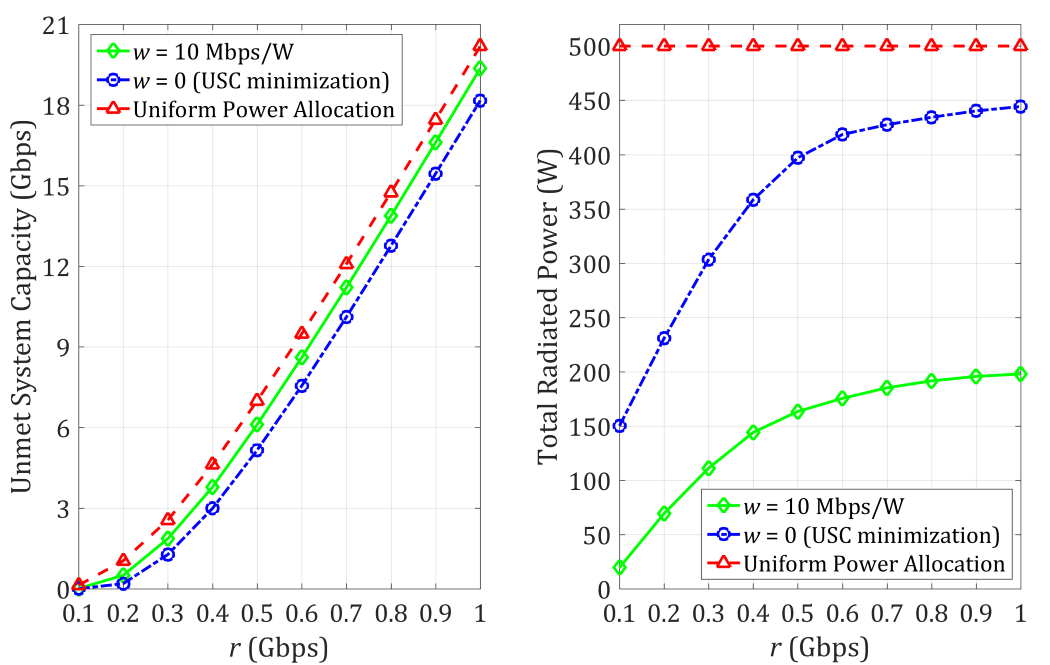} 
\caption[caption]{USC and total radiated power versus the traffic slope.}
\label{Fig2}
\vspace{-2mm}
\end{figure}

\begin{figure}[!t]
\centering
\includegraphics[width=3.35in]{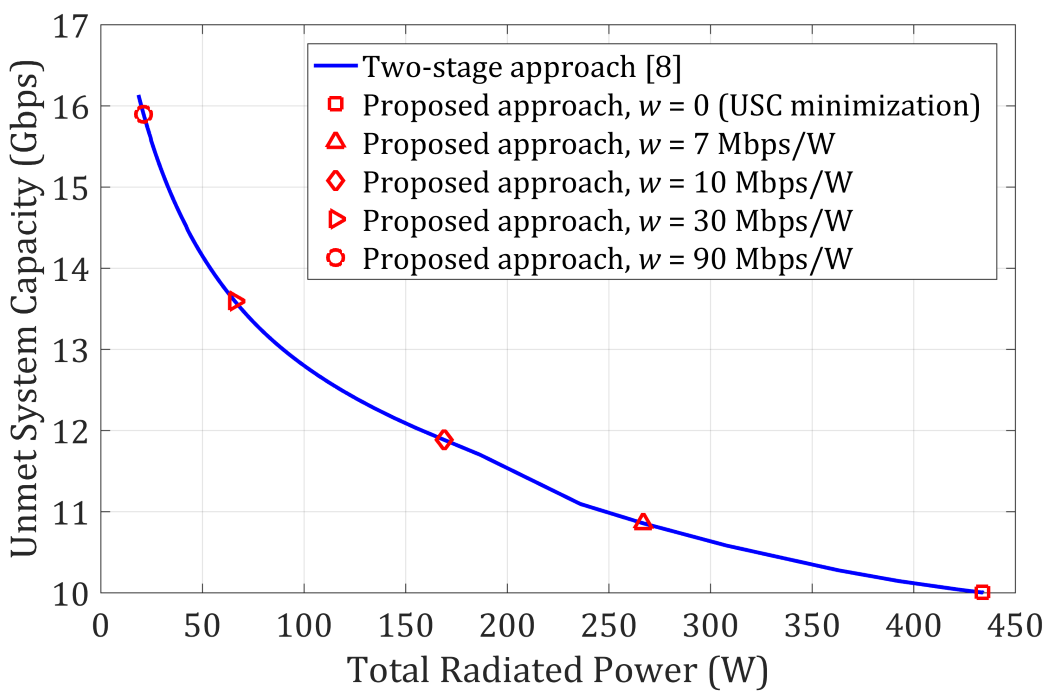} 
\caption{Performance comparison with the two-stage approach \cite{Aravanis} for a particular system configuration with $r = 0.7\,\text{Gbps}$.}
\label{Fig3}
\vspace{-2mm}
\end{figure}

Secondly, Fig. 2 illustrates the USC and total radiated power achieved by the conventional scheme and Algorithm 1 (for two different weights) versus the traffic slope. Although the UPA scheme makes full use of the available power, it has the highest USC. On the other hand, for $w = 0$ (USC minimization) we have the lowest USC using less power than UPA. In addition, the last scheme with $w = 10\,\text{Mbps/W}$ achieves an USC that lies between the other two schemes, but with much less power (high energy savings). This is expected because higher priority is given to the total radiated power as the weight $w$ increases.

Last but not least, Fig. 3 compares the performance of the proposed method with the two-stage approach \cite{Aravanis}. In particular, the 5 operating points attained by the proposed approach belong to the Pareto boundary obtained from \cite{Aravanis}. It has been observed that many values of $w$ achieve operating points on the Pareto boundary, but we only present 5 points for better illustration. Therefore, the proposed method shows similar performance with \cite{Aravanis}. Note that in multi-objective optimization, there is no objectively optimal solution, but only Pareto/subjectively optimal solutions.

In summary, \cite{Aravanis} presents \textit{a posteriori method} where the network designer selects an operating point after the computation/visualization of the Pareto boundary, while this letter introduces \textit{a priori method} where the weight $w$ is specified before any computation, and then a single solution is obtained. Finally, we would like to emphasize that the former approach is appropriate for offline power allocation (no strict limitations on processing time), whereas the latter approach is suitable for online/dynamic power allocation due to its rapid convergence.

\section{Conclusion}
In this letter, we have designed a SCA-based optimization algorithm with high convergence speed, which is suitable for real-time power allocation in MSS with strict computation/processing-time requirements. The proposed multi-objective approach enables network designers to achieve a compromise between the USC and total power consumption. Numerical simulations have also verified the advantage of this approach. Moreover, the complexity of the SCA method, in its general form, has been studied theoretically. 

\appendices
\section*{Appendix \linebreak Successive Convex Approximation Method}
SCA is an iterative method that attains a stationary point of a nonconvex optimization problem by solving a sequence of convex problems \cite{Lanckriet}. Despite the fact that the achieved solution may or may not be globally optimal, this technique has reasonable computational complexity. More specifically, the following theorem is provided, where \textit{all the functions are assumed to be differentiable (and therefore continuous)}.

\vspace{1mm}
\noindent
\textbf{Theorem 1} (\hspace{1sp}\cite{Lanckriet})\textbf{.} \textit{Let $\mathcal{P}$ be a nonconvex minimization problem with objective ${\psi _0}(\boldsymbol{x})$, and nonempty-compact feasible set $D = \{ \boldsymbol{x} \in {\mathbb{R}^n}:\;{\psi _i}(\boldsymbol{x}) \leqslant 0,\; 1 \leqslant i \leqslant m\}$, with $\boldsymbol{x} = [{x_1},{x_2}, \ldots ,{x_n}]$. Moreover, suppose that ${\psi _i}(\boldsymbol{x}) = {u_i}(\boldsymbol{x}) - {v_i}(\boldsymbol{x})$ for $0 \leqslant i \leqslant m$, where ${u_i}(\boldsymbol{x})$ and ${v_i}(\boldsymbol{x})$ are convex functions. Let ${\left\{ {\widetilde {{\mathcal{P}_j}}} \right\}_{j \geqslant 1}}$ be a sequence of convex minimization problems with objective $\widetilde {{\psi _{0,j}}}(\boldsymbol{x},\boldsymbol{x}_{j - 1}^ * )$, compact feasible set ${D_j} = \{ \boldsymbol{x} \in {\mathbb{R}^n}:\; \widetilde {{\psi _{i,j}}}(\boldsymbol{x},\boldsymbol{x}_{j - 1}^ * ) \leqslant 0,\; 1 \leqslant i \leqslant m\}$, and global minimum $\boldsymbol{x}_j^*$ (with $\boldsymbol{x}_0^* \in D$). If $\widetilde {{\psi _{i,j}}}(\boldsymbol{x},\boldsymbol{x}_{j - 1}^ * ) = {u_i}(\boldsymbol{x}) - \widetilde {{v_i}}(\boldsymbol{x},\boldsymbol{x}_{j - 1}^*)$ for $0 \leqslant i \leqslant m$ and $j \geqslant 1$, where $\widetilde {{v_i}}(\boldsymbol{x},\boldsymbol{x}_{j - 1}^ * ) = {v_i}(\boldsymbol{x}_{j - 1}^ * ) + \nabla {v_i}(\boldsymbol{x}_{j - 1}^ * ) \cdot {\left( {\boldsymbol{x} - \boldsymbol{x}_{j - 1}^ * } \right)^T}$, with $\nabla {v_i}(\boldsymbol{x}) = [{{\partial {v_i}(\boldsymbol{x})} \mathord{\left/{\vphantom {{\partial {v_i}(\boldsymbol{x})} {\partial {x_1}}}} \right.\kern-\nulldelimiterspace} {\partial {x_1}}},{{\partial {v_i}(\boldsymbol{x})} \mathord{\left/{\vphantom {{\partial {v_i}(\boldsymbol{x})} {\partial {x_2}}}} \right.\kern-\nulldelimiterspace} {\partial {x_2}}}, \ldots ,{{\partial {v_i}(\boldsymbol{x})} \mathord{\left/{\vphantom {{\partial {v_i}(\boldsymbol{x})} {\partial {x_n}}}} \right. \kern-\nulldelimiterspace} {\partial {x_n}}}]$, then: (a) $\boldsymbol{x}_{j - 1}^ *  \in {D_j} \subseteq D$ and ${\psi _0}(\boldsymbol{x}_j^ * ) \leqslant {\psi _0}(\boldsymbol{x}_{j - 1}^ * )$, $\forall j \geqslant 1$, (b) $\mathop {\lim }\limits_{j \to \infty } {\psi _0}(\boldsymbol{x}_j^ * ) = {\psi _0}(\overset{\lower0.5em\hbox{$\smash{\scriptscriptstyle\frown}$}}{\boldsymbol{x}} ) = L > -\infty$ for all the accumulation/limit points $\overset{\lower0.5em\hbox{$\smash{\scriptscriptstyle\frown}$}}{\boldsymbol{x}}$ of the sequence ${\left\{ {\boldsymbol{x}_j^ * } \right\}_{j \geqslant 0}}$, and (c) assuming suitable constraint qualifications, all the accumulation points $\overset{\lower0.5em\hbox{$\smash{\scriptscriptstyle\frown}$}}{\boldsymbol{x}}$ are stationary points of $\mathcal{P}$ (i.e., satisfy the corresponding Karush-Kuhn-Tucker conditions), and $L = \mathop {\lim }\limits_{j \to \infty } {\psi _0}(\boldsymbol{x}_j^ * ) = {\psi _0}(\hat {\boldsymbol{x}})$, where $\hat {\boldsymbol{x}}$ is some stationary point of $\mathcal{P}$.}
\vspace{1mm}

Taking advantage of the fact that SCA generates a monotonically decreasing sequence of objective values, and using \textit{the property of telescoping sums}: $\sum\nolimits_{l = 1}^{M} {\left( {{a_{l - 1}} - {a_l}} \right)}  = {a_0} - {a_{M}}$ for any integer $M \geqslant 1$, we introduce and prove the following result concerning the complexity of the SCA method. 

\vspace{1mm}
\noindent
\textbf{Theorem 2.} \textit{Suppose that the SCA method terminates when $\left| {{\psi _0}(\boldsymbol{x}_j^ * ) - {\psi _0}(\boldsymbol{x}_{j - 1}^ * )} \right| \leqslant \epsilon \left| {{\psi _0}(\boldsymbol{x}_{j - 1}^ * )} \right|$ for some predefined tolerance $\epsilon > 0$, and ${\psi _0}({\boldsymbol{x}^ * }) > 0$, where $\boldsymbol{x}^ *$ is a global minimum of $\mathcal{P}$. Then, the complexity of the SCA method is $\mathcal{O}\left( {\left( {{\xi \mathord{\left/{\vphantom {\xi }} \right.\kern-\nulldelimiterspace} } {\epsilon}} \right)h(n,m)} \right)$, where $\xi  = {{{\psi _0}(\boldsymbol{x}_0^ * )} \mathord{\left/{\vphantom {{{\psi _0}(\boldsymbol{x}_0^ * )} {{\psi _0}({\boldsymbol{x}^ * })}}} \right.\kern-\nulldelimiterspace} {{\psi _0}({\boldsymbol{x}^ * })}} \geqslant 1$ and $h(n,m)$ is the complexity of each convex optimization problem which is a polynomial function of the number of variables and constraints ($n$ and $m$, respectively).}
\vspace{1mm}

\textit{Proof:} According to Theorem 1, it holds that ${\psi _0}({\boldsymbol{x}}_0^ * ) \geqslant {\psi _0}({\boldsymbol{x}}_{j - 1}^ * ) \geqslant {\psi _0}({\boldsymbol{x}}_j^ * ) \geqslant {\psi _0}({\boldsymbol{x}^ * }) > 0$, $\forall j \geqslant 1$. As concerns the number of iterations until convergence, if we denote by $\nu$ the smallest integer such that $\left| {{\psi _0}(\boldsymbol{x}_\nu^ * ) - {\psi _0}(\boldsymbol{x}_{\nu - 1}^ * )} \right| \leqslant \epsilon \left| {{\psi _0}(\boldsymbol{x}_{\nu - 1}^ * )} \right|$ $\Leftrightarrow$ \linebreak ${\psi _0}(\boldsymbol{x}_{\nu - 1}^ * ) - {\psi _0}(\boldsymbol{x}_\nu^ * ) \leqslant \epsilon \, {\psi _0}(\boldsymbol{x}_{\nu - 1}^ * )$, then for all integers less than $\nu$ the last inequality does not hold: ${\psi _0}(\boldsymbol{x}_{l - 1}^ * ) - {\psi _0}(\boldsymbol{x}_l^ * ) > \epsilon \, {\psi _0}(\boldsymbol{x}_{l - 1}^ * ) \geqslant \epsilon \, {\psi _0}({\boldsymbol{x}^ * })$ $\Rightarrow$ ${\psi _0}(\boldsymbol{x}_{l - 1}^ * ) - {\psi _0}(\boldsymbol{x}_l^ * ) > \epsilon \, {\psi _0}({\boldsymbol{x}^ * })$, $\forall l \in \left\{ {1,2, \ldots ,\nu - 1} \right\}$. By summing from $1$ to $\nu - 1$, we obtain $\sum\nolimits_{l = 1}^{\nu - 1} {\left( {{\psi _0}(\boldsymbol{x}_{l - 1}^ * ) - {\psi _0}(\boldsymbol{x}_l^ * )} \right)}  > \sum\nolimits_{l = 1}^{\nu - 1} \epsilon \, {{\psi _0}({\boldsymbol{x}^ * })}$ $\Rightarrow$ ${\psi _0}(\boldsymbol{x}_0^ * ) - {\psi _0}(\boldsymbol{x}_{\nu - 1}^ * ) > \left( {\nu - 1} \right) \epsilon \, {\psi _0}({\boldsymbol{x}^ * })$. Since ${\psi _0}(\boldsymbol{x}_{\nu - 1}^ * ) \geqslant {\psi _0}({\boldsymbol{x}^ * })$, we get $\left( {\nu - 1} \right) \epsilon \, {\psi _0}({\boldsymbol{x}^ * }) < {\psi _0}(\boldsymbol{x}_0^ * ) - {\psi _0}({\boldsymbol{x}^ * })$, and therefore $\nu < 1 + {{\left( {\xi  - 1} \right)} \mathord{\left/{\vphantom {{\left( {\xi  - 1} \right)} }} \right.\kern-\nulldelimiterspace} {\epsilon}} < 1 + {{ {\xi} } \mathord{\left/{\vphantom {{ {\xi} } }} \right.\kern-\nulldelimiterspace} {\epsilon}} = \mathcal{O}\left( {{\xi \mathord{\left/{\vphantom {\xi }} \right.\kern-\nulldelimiterspace} {\epsilon}}} \right)$. Hence, the SCA method requires $\mathcal{O}\left( {{\xi\mathord{\left/{\vphantom {\xi }} \right.\kern-\nulldelimiterspace} {\epsilon}}} \right)$ iterations to converge. Moreover, each convex optimization problem can be globally solved with polynomial complexity in the number of variables and constraints \cite{Boyd}, and thus Theorem 2 follows directly. \hfill  $\blacksquare$


\end{document}